\documentclass[journal=jacsat,manuscript=communication,articletitle=true,layout=twocolumn,]{achemso}
\usepackage[usenames,dvipsnames]{xcolor}
\usepackage[bookmarks=false,colorlinks]{hyperref}
\hypersetup{
    linkcolor=black, 
	citecolor=black, 
	filecolor=black, 
	urlcolor=black,  
} 
\usepackage{amsmath,amssymb}
\usepackage{mdwlist} 
\usepackage{booktabs}
\usepackage{multirow}
\usepackage{amsmath}
\usepackage{adjustbox}

\definecolor{col1}{rgb}{0.0, 0.46, 0.8}
\definecolor{col2}{rgb}{0.9, 0.0, 0.30}
\definecolor{col3}{rgb}{0,0,154}

\usepackage{lineno}

\usepackage[draft]{changes} 
\usepackage[normalem]{ulem}

\title{Thermal stabilities Landscape of A$_2$BB$^{\prime}$O$_6$ compounds}

\author{Yateng Wang}
\affiliation{Beijing Advanced Innovation Center for Materials Genome Engineering, University of Science and Technology Beijing, Beijing, 100083 China}
\alsoaffiliation{School of Mathematics and Physics, University of Science and Technology Beijing, Beijing 100083, China}

\author{Bianca Baldassarri}
\affiliation{Department of Materials Science and Engineering, Northwestern University, Evanston, Illinois 60208, United States}

\author{Jiahong Shen}
\affiliation{Department of Materials Science and Engineering, Northwestern University, Evanston, Illinois 60208, United States}

\author{Jiangang He}
\email{jghe2021@ustb.edu.cn}
\affiliation{Beijing Advanced Innovation Center for Materials Genome Engineering, University of Science and Technology Beijing, Beijing, 100083 China}
\alsoaffiliation{School of Mathematics and Physics, University of Science and Technology Beijing, Beijing 100083, China}

\author{Chris Wolverton}
\email{c-wolverton@northwestern.edu}
\affiliation[Northwestern University]{Department of Materials Science and Engineering, Northwestern University, Evanston, Illinois 60208, United States}

\begin{document}
\begin{abstract}
Perovskite oxides have been extensively studied for their wide range of compositions and structures, as well as their valuable properties for various applications. Expanding from single perovskite ABO$_3$ to double perovskite $A_2BB^{\prime}$O$_6$ significantly enhances the ability to tailor specific physical and chemical properties. However, the vast number of potential compositions of $A_2BB^{\prime}$O$_6$ makes it impractical to explore them all experimentally. In this study, we conducted high-throughput calculations to systematically investigate the structures and stabilities of 4,900 $A_2BB^{\prime}$O$_6$ compositions (with $A$ = Ca, Sr, Ba, and La; $B$ and $B^{\prime}$ representing metal elements) through over 42,000 density functional theory (DFT) calculations. Our analysis lead to the discovery of more than 1,500 new synthesizable $A_2BB^{\prime}$O$_6$ compounds, with over 1,100 of them exhibiting double perovskite structures, predominantly in the $P2_1/c$ space group. By leveraging the high-throughput dataset, we developed machine learning models that achieved mean absolute errors of 0.0444 and 0.0330 eV/atom for formation energy and decomposition energy, respectively. Using these models, we identified 803 stable or metastable compositions beyond the chemical space covered in our initial calculations, with 612 of them having DFT-validated decomposition energies below 0.1 eV/atom, resulting in a success rate of 76.2 \%. This study delineates the stability landscape of $A_2BB^{\prime}$O$_6$ compounds and offers new insights for the exploration of these materials.
\end{abstract}

\maketitle

\noindent $\blacksquare$ \textbf{Introduction}
Perovskite oxides $AB$O$_3$ ($A$= alkaline-earth and rare-earth elements; $B$ = transition metal elements) have attracted significant research interest due to their unique properties,\cite{tejuca1992properties,arul2020revolution} such as metal-insulator transition,\cite{fath1999spatially} ferroelectrics and multiferroic behavior,\cite{cohen1992origin,doi:10.1126/science.1080615} half-metallic spin-polarized electrical conductivity,\cite{7160747} magnetocaloric effect,\cite{10.1063/1.1405836} colossal magnetoresistance,\cite{ramirez1997colossal} superconductivity,\cite{maeno1994superconductivity} among others. These exceptional properties render perovskite oxides highly promising for a diverse array of applications, encompassing magnetic refrigeration,\cite{10.1063/1.1405836} magnetic random access memory,\cite{7160747} spintronic devices, photocatalysis,\cite{doi:10.1021/cr1002326} solid oxide fuel cells,\cite{doi:10.1021/cr020724o,grimaud2013double} energy storage,\cite{cheng2012metal} solar thermal chemical water splitting,\cite{C8EE01989D} thermoelectricity,\cite{MUTA2005306} and biomedical applications.\cite{arul2020revolution} Additionally, the capacity to manipulate and regulate these properties positions perovskite as a captivating field of study with the potential for substantial technological advancements. This exceptional diversity and adjustable properties originate from a wide range of compositions, encompassing nearly all the elements in the periodic table, and the flexible structures of the corner-sharing octahedral. The perovskite structure was originally named after Russian geologist Perovski.\cite{rose1839beschreibung} It is now extended to stand for a group of the compounds that contain the chemical formula $AB$O$_3$ and are characterized by corner-sharing $B$O$_6$ octahedral.

Although the number of potential $AB$O$_3$ compounds and the derivations ($A_2BB^{\prime}$O$_6$, $AA^{\prime}B_2$O$_6$, and $AA^{\prime}BB^{\prime}$O$_6$) exceed 10$^5$,\cite{C8EE01574K} only around 1000 of them have been synthesized experimentally so far.\cite{vasala2015a2b} Exploring such an extensive chemical phase space through experimentation is exceedingly challenging due to the time-consuming nature of experimental synthesis and the unique synthesis conditions required for each compound. Therefore, a revolution has transpired in materials design and discovery, as the simulation of experimental synthesis and characterization is now achievable through high-throughput density functional theory (DFT) calculations.\cite{gautier2015prediction,tabor2018accelerating,emery2016high,he2022computationally} Thermodynamic stability is a key parameter that broadly governs whether a material is expected to be synthesizable (or whether it may decompose under certain environment conditions) and can be computed easily using DFT.\cite{r2007first} As a result, computational materials discovery through thermodynamic computation has been employed in various categories of inorganic materials,\cite{Alberi_2019} such as Heusler compounds,\cite{gautier2015prediction,he2022computationally} perovskites,\cite{emery2016high,yang2022high,doi:10.1021/acs.chemmater.9b00116,JIANG2021351,C9TA01456J,wang2021high,C5TC04172D}  garnets,\cite{schmidt2023machine} delafossite,\cite{shi2017high} mixed anion compounds,\cite{doi:10.1021/acs.chemmater.0c01902,shen2021high} quaternary chalcogenides,\cite{pal2022scale} etc., with applications ranging from lithium-ion batteries,\cite{kirklin2013high,aykol2016high,MUY2019270} to transparent conductors,\cite{hautier2013identification,C7TC05311H,C9TA01456J,shi2017high} optoelectronic devices,\cite{PhysRevMaterials.3.044602,petousis2017high,C5TC04172D} electrides,\cite{doi:10.1021/acs.chemmater.8b02526} catalysis,\cite{C8EE01574K} photovoltaics,\cite{C8EE01574K,https://doi.org/10.1002/adfm.201804354} thermoelectrics,\cite{pal2022scale,doi:10.1021/acs.chemmater.0c01902} and to solar thermochemical water splitting.\cite{emery2016high,PhysRevMaterials.7.065403} Emery \textit{et al.} computationally screened 5329 compositions of $AB$O$_3$ compounds considering four different perovskite structures (cubic, rhombohedral, tetragonal, and orthorhombic) and found 383 stable $AB$O$_3$ compounds.\cite{emery2016high} Schmidt \textit{et al.} screened 250,000 cubic perovskite and antiperovskite compositions by employing DFT calculations and discovered 500 new stable compounds.\cite{doi:10.1021/acs.chemmater.7b00156} K\"orbel \textit{et al.} performed high throughput DFT screening on $ABX_3$ compounds in a chemical space containing 64 elements ($A$ and $B$ are 55 metal elements and $X$ = nitrogen, oxygen, chalcogenides, and halides).\cite{C5TC04172D} They found 199 stable and metastable $ABX_3$ compounds with various properties from 32,000 candidates. Zhao \textit{et al.} investigated lead-free inorganic halide double perovskites computationally toward photovoltaic application.\cite{doi:10.1021/jacs.6b09645} Eleven promising photovalitic materials with high thermodynamic stability, suitable band gaps, small carrier effective masses, and low exciton binding energies were discovered by screening 64 candidate compositions.

Nevertheless, high-throughput computational screening requires substantial computational resources and has a relatively low success rate. With the advancement of data science, the use of machine learning (ML) and machine-learning-assisted high-throughput screening has become increasingly prevalent in the field of materials science, leading to accelerated materials discovery.\cite{ward2016general,schmidt2019recent} For example, Faber \textit{et al.} screened 2 million elpasolite ($ABC_2D_6$ chemical formula with $Fm\bar{3}m$ space group) compounds using a ML model trained with ten thousand DFT formation energies.\cite{faber2016machine} The accuracy of this model reaches 0.1 eV/atom and 90 stable compounds were identified. Li \textit{et al.} trained an ML model of thermodynamic stability using the convex hull data of 1,929 perovskite oxides stored in the Materials Project.\cite{LI2018454} The RMSE (root mean squared error) of 28.5 meV/atom and MAE (mean absolute error) of 16.7 meV/atom were accomplished by using the selected 70 features, which are based on elemental properties. After screening 21,316 compositions with the established ML model, 764 new perovskites were predicted and 98 of them were validated to be stable by DFT calculations. Kim \textit{et al.} developed a regression model on formation energies of $A_2BB^\prime$O$_6$ double perovskites with R$^2$ (coefficient of determination) of 0.98 and RMSE of 0.175 eV/atom and a classification model of the stability with F1-score of 0.771 by means of training the formation energies of 1,052 double perovskites in the Materials Project database.\cite{10.1063/1.4812323,kim2021synthesizable} These two models are employed to discover synthesizable $A_2BB^{\prime}$O$_6$ compounds from 11,763 candidates. The DFT validated accuracy of these two models are 0.646 and 0.733 for the convex hull energy and F1-score, respectively. A ML model based on crystal graph and convolutional neural networks (CGCNN) established by Xie and Grossman was trained with DFT-calculated data in the Materials Project to predict new perovskites and 228 new compounds were predicted to be synthesisable.\cite{PhysRevLett.120.145301} Li \textit{et al.} developed a deep transfer learning approach based on the formation energies of spinel oxides and predicted 1,314 thermodynamically stable perovskite oxides by training 5,329 spinel oxides and 855 perovskite oxides.\cite{li2023center}

Despite its advantages, machine learning necessitates a substantial volume of data for model training, which has to be either carefully extracted from experiments/literature or generated through massive DFT calculations when the data is unavailable in established databases. Therefore, a nature marriage between high-throughput DFT calculations and machine learning can significantly accelerate material discovery procedure, allowing to access materials in the compositional spaces that are too large even for high-throughput DFT calculations. Schmidt \textit{et al.} developed a ML model utilizing the formation energies of 249,984 cubic $ABX_3$ perovskites computed by high-throughput DFT calculations.\cite{doi:10.1021/acs.chemmater.7b00156} Their ML model can reproduce the formation energy data with errors around 0.1 eV/atom.

In this work, we performed a high throughput DFT calculations on $A_2BB^{\prime}$O$_6$ ($A$= Ca, Sr, Ba, and La; $B$ and $B^{\prime}$ are metal elements) compositions with 10 common (5 perovskite and 5 non-perovskite) crystal structures. We found 2022 stable/metastable compounds and 1785 of them favor perovskite structures. Our statistic analysis based on ML model shows that the formation energy of $A_2BB^{\prime}$O$_6$ compounds is strongly correlated to element properties such as the number of unfilled electrons in the outermost layer of an atom, the number of valence electrons in an atom, the thermodynamic properties of elements, and weakly correlated to the known features such as tolerant factors. With the established ML model, we screened the $A_2BB^{\prime}$O$_6$ compositions out of our chemical space and discovered 612 stable/metastable $A_2BB^{\prime}$O$_6$ compounds by verifying our prediction using the OQMD database.

\vspace{0.5cm}
\noindent $\blacksquare$ \textbf{Methodology}

\noindent
\textbf{DFT parameters.} In this work, all DFT calculations are performed using the Vienna \textit{ab initio} Simulation Package (VASP).\cite{vasp1,vasp2} The Perdew-Burke-Ernzerhof (PBE) version of the exchange-correlation functional,\cite{PBE} projector-augmented wave (PAW) method,\cite{PAW1, PAW2} and the plane wave basis sets with energy cutoff of 520 eV were employed for structure relaxation and static calculations. The $\Gamma$-centered $k$-point grids with a density of 8000 $k$-points per reciprocal atom (KPPRA) were used. The compounds containing the elements with unfilled $d$-shells were carried out with spin-polarized calculations in a ferromagnetic spin configuration, and the magnetic degree of freedom allowed to relax to self-consistency within a primitive unit cell. Other parameters, such as Hubbard U values,\cite{dudarev1998electron} are consistent with the OQMD setting.\cite{kirklin2015open}

\begin{figure}[htp!]
	\setlength{\unitlength}{1cm}
	\includegraphics[width=1.0\linewidth,angle=0]{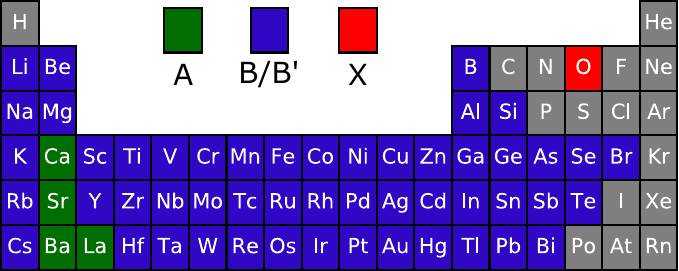}	
	\caption{$A$, $B$, and $B^{\prime}$-site elements used in the $A_2BB{^\prime}X_6$ compounds of this work.}
	\label{chemsapce}
\end{figure}

\vspace{0.2cm}
\noindent
\textbf{Chemical space.} In this work, we considered 4 commonly observed $A$-site elements (Ca, Sr, Ba, and La) in perovskites and 50 metal elements for $B$ and $B^{\prime}$ sites, see the element list in Figure~\ref{chemsapce}. Perovskite is the most common structure type of $A_2BB^{\prime}$O$_6$ and its $A$-site has high coordination number (e.g., 12 for cubic perovskite), which requires the cation with large radius, \textit{i.e.}, Pauling's first rule.\cite{doi:10.1021/ja01379a006} We considered all the combination of $B$ and $B^{\prime}$ elements, which generates 4,900 (4 $\times$ $C_{50}^2$) $A_2BB^{\prime}$O$_6$ compositions. Rare earth elements are excluded from $B$ and $B^{\prime}$ sites because they are not common at the $B$-site of perovskite.\cite{vasala2015a2b}

\noindent \textbf{Thermodynamic stability at T=0 K.} The thermodynamic stability of a compound at 0 K can be estimated by its energetic difference from the convex hull of the formation energies in a given chemical space, which is the energy difference between the formation energy of the compound and that of the lowest linear combination of the phases at the same composition. Previous studies indicated that the distance to the convex hull of formation energies (commonly referred to as ``convex hull distance'' or ``stability'') serves as a reliable metric for assessing the likelihood of experimental synthesizability: a convex hull distance of zero or close to zero is a strong indicator of successful experimental synthesis.\cite{PhysRevB.95.024411,PhysRevB.98.094410,emery2016high,doi:10.1021/acs.chemmater.0c01902} All unique and ordered phases in the OQMD (\href{http://oqmd.org/}{\textcolor{blue}{www.oqmd.org}}), a total of over 1,000,000 phases (as of June 2021), including $\sim$40,000 known compounds from the Inorganic Crystal Structure Database (ICSD)\cite{ICSD1, ICSD2} and $\sim$600,000 hypothetical compounds based on common binary, ternary, and quaternary structure types,\cite{OQMD} were used for all convex hull constructions. The hull distance ($\Delta{E_{\rm{h}}}$) of an $A_2BB^{\prime}$O$_6$ compound is defined as the energy difference between its formation energy ($\Delta{E}$) and the lowest linear combination of the formation energies of the phases in the $A$-$B$-$B^{\prime}$-O phase space with an overall composition of $A_2BB^{\prime}$O$_6$. These phases are referred to as ``competing phases'' in this work.

\vspace{0.2cm}
\noindent \textbf{Crystal structure determination.} Crystal structure plays a crucial role for a compound as it determines its properties. Although many compounds can have several polymorphs at ambient condition, the lowest energy structure is generally the most common form of a compound. Moreover, the lowest energy structure is particularly useful to precisely evaluate thermal stability of a composition. However, it is very computationally expensive to determine the lowest-energy crystal structure of a composition containing multiple elements. Although perovskite is the most common structure of $A_2BB^{\prime}$O$_6$ chemical formula, many non-perovskite structures do exist and have lower energy than perovskite. Furthermore, perovskite has many distorted structures, which have lower energies than the cubic one. Owing to the high computational cost of convention crystal structure prediction approaches, such as \texttt{USPEX}~\cite{GLASS2006713} and \texttt{CALYPSO}~\cite{PhysRevB.82.094116,wang2012calypso}, it is too expensive to perform convention crystal structure search for so many compositions. Therefore, we performed high-throughput structure search using the prototype structures approach. Since double perovskite structures are the most commonly observed structures in $A_2BB^{\prime}$O$_6$ and the distortion of perovskite is very common and is strongly correlated to $A$ and $B$-site cation size mismatch, we considered several double perovskite structures with different octahedral distortion to cover a wide range of radius mismatch between $A$ and $B$ ($B^{\prime}$)-site cations, which is often characterized by the Goldschmidt tolerance factor ($\tau$).\cite{goldschmidt1926gesetze} Here, we distinguish the $B$/$B^{\prime}$ ordered $A_2BB^{\prime}$O$_6$ perovskite from $B$/$B^{\prime}$ disordered one. If the $B$-site cations are disordered, the compound is labeled as doped or disordered single perovskite, instead of double perovskite. There are three common pattern of forming $B$/$B^{\prime}$ ordered double perovskite: rock-salt, columnar, and layered.\cite{B926757C} The know $A_2BB^{\prime}$O$_6$ compounds summarized by Vasala and Karppinen\cite{vasala2015a2b} indicates 58.1 \% of $A_2BB^{\prime}$O$_6$ perovskites are rock-salt ordered, while only 0.5\% of them are layered. The columnar ordering is even more uncommon. Our further analysis on rock-salt ordered $A_2BB^{\prime}$O$_6$ (also known as Elpasolite structure, after the mineral K$_2$NaAlF$_6$) compounds show that 54.3 \%, 30.2 \%, 8.7 \%, and 3.4 \% of them are monoclinic, cubic, tetragonal, and orthorhombic, respectively. Among monoclinic, $P2_1/c$ is the dominant space group, which can be viewed as the $B$/$B^{\prime}$ rock-salt ordered $Pnma$ of single perovskite. Similar to $Pnma$, which is the most common structure of single perovskite $AB$O$_3$,\cite{Lufaso:br0106} $P2_1/c$ is a distorted perovskite formed by octahedral rotation and titling with respect to the cubic structure ($Fm\bar{3}m$).\cite{Howardta5000} If only the rotation ($X_3^{+}$ in the language of irreducible representation, similar to $R_4^{+}$ of perovskite) is involved, the subgroups of $Fm\bar{3}m$ are $I4/m$ ($a^0a^0c^-$, Galzer notation\cite{glazer1972classification}), $C2/m$ ($a^0b^-b^-$), and $R\bar{3}$ ($a^-a^-a^-$). If only the tilting ($\Gamma_4^+$, similar to $M_3^{+}$ of perovskite $AB$O$_3$) is involved, the subgroups of $Fm\bar{3}m$ are $Pn\bar{3}$ ($a^+a^+a^+$), $P4_2/nnm$ ($a^0b^+b^+$), and $P4/mnc$ ($a^0a^0c^+$). If both rotation and tilting are involved, 6 subgroups may form. But the most common structures are $P2_1/c$ ($a^+b^-b^-$) and $C2/c$ ($a^+b^+c^-$). However, rotation usually gains more energy than tilting,\cite{PhysRevB.89.045104,doi:10.1021/acs.chemmater.6b03486} therefore the most commonly observed double perovskite structures are $P2_1/c$, $C2/m$, $R\bar{3}$, $I4/m$, and $Fm\bar{3}m$. Also, the structure distortion is strongly dependent on temperature and phase transitions are commonly observed in many double perovskites. $Fm\bar{3}m$ is the high-temperature phase of many double perovskite compounds, though the rotation and tilting distorted structures are commonly adopted at low temperature. Moreover, Cs$_2$Au$_2$Cl$_6$-type ($I4/mmm$) double perovskite structure is formed through charge disproportion of 2Au$^{2+}$ $\rightarrow$ Au$^{+}$ + Au$^{3+}$, which is similar to BaBiO$_3$.\cite{doi:10.1021/acs.chemmater.6b03486} Ba$_2$InCuO$_6$-type ($P\bar{3}m1$) and Ba$_2$NiTeO$_6$-type ($R\bar{3}m$) are the commonly observed non-perovskite structures with large $A$-site radius, while Li$_2$SnTeO$_6$-type ($Pnn2$) and Ni$_3$TeO$_6$-type ($R3$) are more common in the non-perovskite $A_2BB^{\prime}$O$_6$ compounds with small $A$-site radius.\cite{su2019predicted}

Therefore, we use the $B$/$B^{\prime}$ rock-salt-ordered cubic double perovskite ($Fm\bar{3}m$) as the first prototype structure to perform the screening of 4,900 compositions because this structure (high symmetry and less free parameters, only contains 10 atom per primitive unit cell) requires a smaller amount of computational resources. Then, for the 2815 compositions that are within 200 meV/atom from the convex hull, we perform further prototype structure screening with the other 9 structures, including 5 perovskite with different distortion ($P2_1/c$, $C2/m$, $I4/m$, $R\bar{3}$, $I4/mmm$) and 4 non-perovskite (Ba$_2$InCuO$_6$-type, $P\bar{3}m1$; Ba$_2$NiTeO$_6$-type, $R\bar{3}m$; Li$_2$SnTeO$_6$-type, $Pnn2$; Ni$_3$TeO$_6$-type, $R3$) structures, see Figure \textcolor{blue}{S1} of the supporting information. As mentioned above, these structures cover a wide range of radius mismatch between $A$ and $B$-site cations. This procedure can significantly reduce the computational cost but may miss the compounds with the ground state structure that is more than 200 meV/atom lower than $Fm\bar{3}m$. However, this chance should be quite low since 200 meV/atom is a pretty large number and $Fm\bar{3}m$ is a reasonable guess for most compounds.

\vspace{0.5cm}
\noindent \textbf{Machine learning model.} Our DFT computed formation energies and decomposition energies of these compounds are used to train machine learning modes to unveil the underlying correlation between formation energy/stability and the fundamental properties of constituent elements and to accelerate new $A_2BB^{\prime}$O$_6$ compounds discovery. 443 element-based features are generated using Matminer,\cite{WARD201860} including element information such as Mendeleev number, occupation of $d$-orbitals, electronegativity, the properties of elemental bulk (melting temperature, covalent radius, \textit{etc.}), and the properties of forming common compounds (oxidation state). Additionally, we adopted 146 features of perovskite from literature.\cite{WANG2022110899} 

Since both formation energy and decomposition energy are continuous numbers, 23 regression models are tested in this study. All the algorithm except XGBoost, which is from XGBoost package,\cite{Chen:2016:XST:2939672.2939785} are adopted from scikit-learn package.\cite{pedregosa2011scikit} The hyper-parameters are determined using a 5-fold cross-validation with a view to maximize accuracy while minimize the standard deviation on unseen data. We use recursive feature elimination (RFE) and Spearman rank to perform feature selection, we use greedy search to find the optimal parameters. Please see Table \textcolor{blue}{S1} of the supporting information for more details. The data is randomly split into 80\% for training and 20 \% for cross-validation for 10 times. The averaged R$^2$ score, RMSE, and MAE are used to evaluate the performance of models.

\begin{figure*}[th!]
	\setlength{\unitlength}{1cm}
	\includegraphics[width=1.0\linewidth,angle=0]{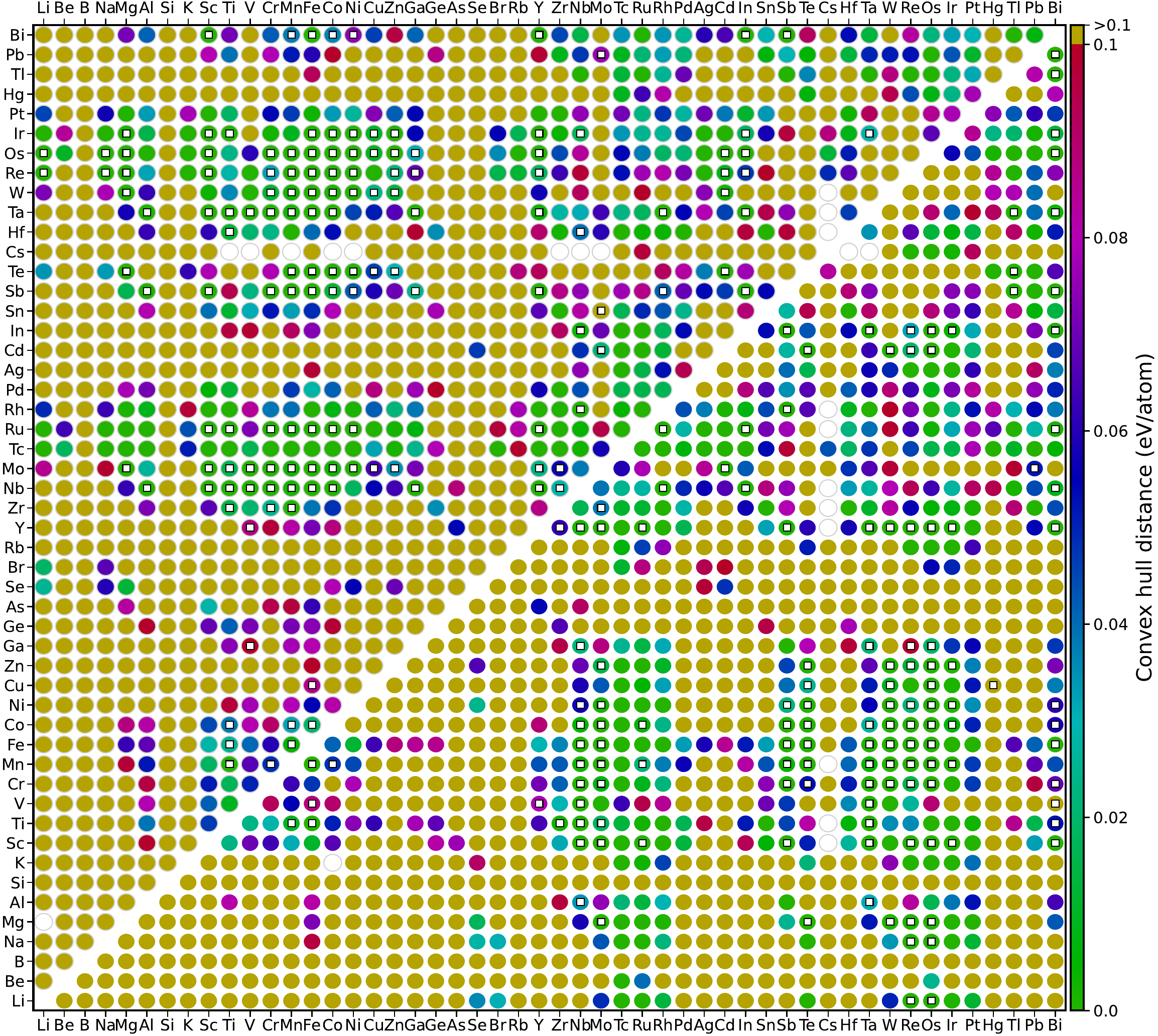}	
	\caption{Heat map of the convex hull distance of Ba$_2BB^{\prime}$O$_6$ (upper triangle) and Sa$_2BB^{\prime}$O$_6$ (lower triangle) compounds. White unfilled circles indicate that the calculations of the corresponding compound fails. Small empty squares inside the larger circle indicate that the compound has been reported in the ICSD or literature.}
	\label{basr}
\end{figure*}

\begin{figure*}[htp!]
	\setlength{\unitlength}{1cm}
	\includegraphics[width=1.0\linewidth,angle=0]{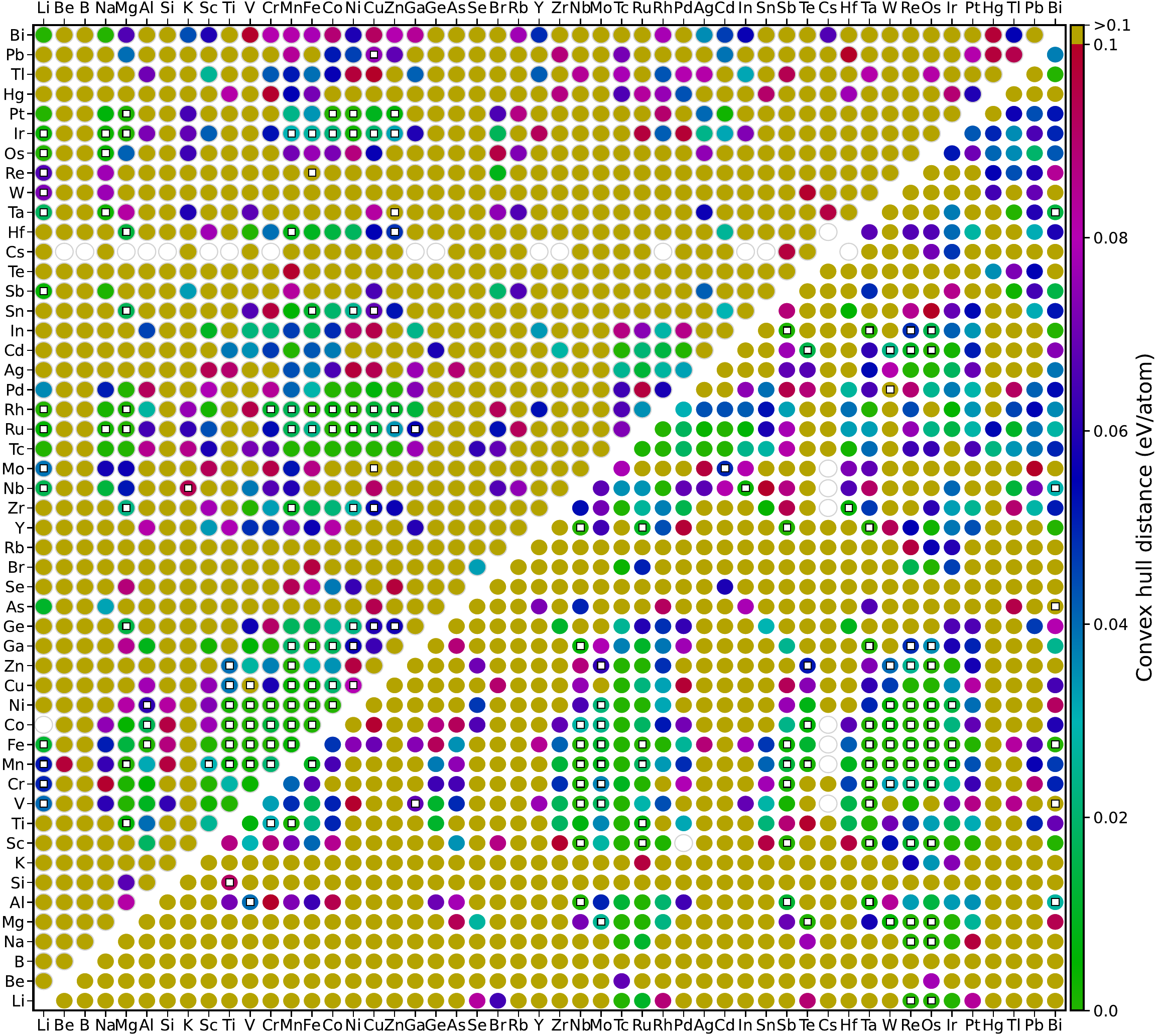}	
	\caption{Heat map of the convex hull distance of La$_2BB^{\prime}$O$_6$ (upper triangle) and Ca$_2BB^{\prime}$O$_6$ (lower triangle) compounds. White unfilled circles indicate that the calculations of the corresponding compound fails. Small empty squares inside the larger circle indicate that the compound has been reported in the ICSD or literature.}
	\label{cala}
\end{figure*}

\vspace{0.5cm}
\noindent $\blacksquare$ \textbf{Results and Discussion}\\
\noindent \textbf{Thermodynamic stability at 0~K.} Thermodynamic stability forms the basis of material properties. It is particularly critical when predicting as-yet synthesized materials, \textit{i.e.}, computational materials discovery. Ideally, the thermodynamic stability should be evaluated at room temperature or the conditions of service. However, the entropy contribution to free energy, especially the vibrational entropy of solids, is computationally costly for a large set of materials. Also, the ambient pressure (10$^{-4}$ GPa) is approximately to zero for solids, therefore the contribution of work (product of pressure and volume) is negligible to free energy at ambient condition. Therefore, we ignore the contribution from entropy and work terms in the Gibbs free energy, and only use zero-kelvin energy to estimate thermodynamic stability of solids.\cite{doi:10.1021/acs.chemmater.0c01902} Following the DFT screening for 4,900 compositions with $Fm\bar{3}m$ structure, our DFT screening with 9 prototype structures for the compositions that are within 200 meV/atom above the convex hull generate 420 stable and 1,602 metastable (convex hull distance is less than 100 meV/atom) compounds, which is much larger than the known $A_2BB^{\prime}$O$_6$ compounds reported in the ICSD and literature, see Table~\ref{stablecomp}.\cite{vasala2015a2b,ICSD1,ICSD2} Note the compounds reported in the ICSD and literature include both $B$/$B^{\prime}$ ordered double perovskites, doped perovskite, and non-perovskite, whereas in this work we only studied $B$/$B^{\prime}$ ordered $A_2BB{^\prime}$O$_6$ compounds. The configurational entropy of $B$/$B^{\prime}$ disorder can lower the free energy of the $B$/$B^{\prime}$ disordered compounds at elevated temperatures, and therefore the observed $A_2BB{^\prime}$O$_6$ structure may be $B$/$B^{\prime}$ disordered at finite temperature if the energy gained by ordering $B$/$B^{\prime}$ can be canceled by configurational entropy at certain temperatures. The distribution of the convex hull distance of all these $A_2BB{^\prime}$O$_6$ compounds are depicted in Figure~\ref{basr} and \ref{cala} for Ba/Sr (Ba: upper triangle; Sr: lower triangle) and Ca/La (Ca: upper triangle; La: lower triangle), respectively. The compounds that have been experimentally reported are indicated by smaller open squares within the color-coded circles. We can see that most of the reported compounds exhibit green color, \textit{i.e.}, on the convex hull or very close to the hull. This is consistent with previous observations in many other compounds.\cite{emery2016high,doi:10.1021/acs.chemmater.0c01902,pal2021accelerated,doi:10.1021/acs.chemmater.1c02294}

\begin{figure}[th!]
	\setlength{\unitlength}{1cm}
	\includegraphics[width=1.0\linewidth,angle=0]{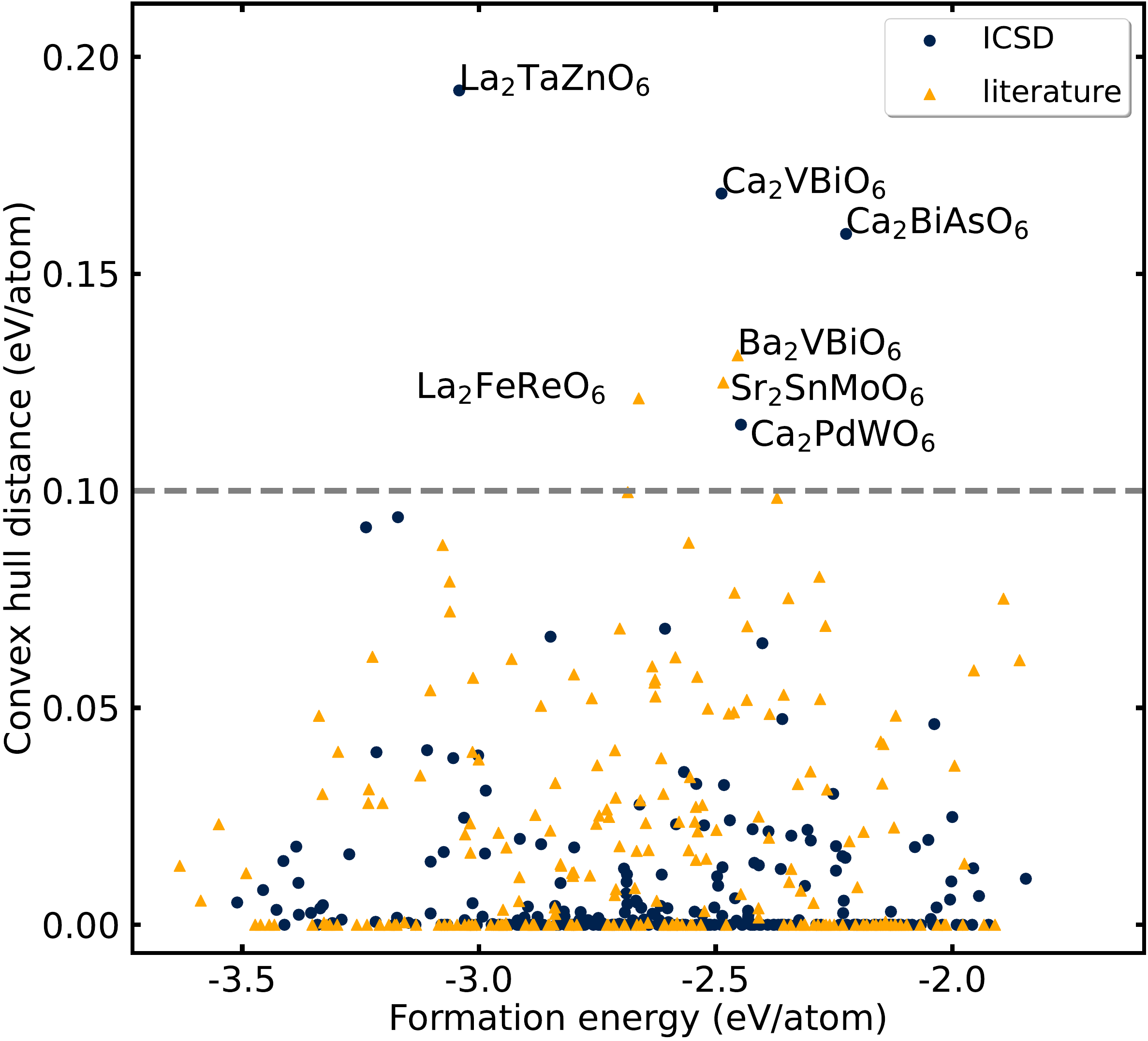}	
	\caption{DFT-calculated formation energy vs convex hull distance of all the reported $A_2BB^{\prime}$O$_6$ compounds in the chemical space defined in this work. A hull distance E$_{\rm stable}$ = 0 (on the convex hull) indicates the compound is thermodynamically stable at 0 K.}
	\label{stableexpt}
\end{figure}

Note the $A_2BB{^\prime}$O$_6$ compounds reported in literature haven't been included in the OQMD prior to our high throughput DFT calculations, the ``re-discovery'' of these compounds confirms our metric of stability for these compounds, which is more reliable than formation energy (all the formation energies of the studied compounds are negative but the majority of them are unstable) because it includes phase competition, which is more important in multi-component systems. Comparing with experiments, there are only 7 exceptions, as indicated in Figure~\ref{stableexpt}: Ba$_2$VBiO$_6$, Sr$_2$MoSnO$_6$, Ca$_2$PdWO$_6$, Ca$_2$AsBiO$_6$, La$_2$FeRhO$_6$, La$_2$VCuO$_6$, La$_2$ZnTaO$_6$. The lowest-energy structure of Ba$_2$BiVO$_6$ in our calculations is $R3$, which is 131 meV/atom above the hull, while the experimentally observed structure is pretty complex: a non-perovskite structure with $P2_1/a$ space group, large unit cell (a=24.640 \AA, b=7.73 \AA, c=5.63 \AA), and octahedrally coordinated Bi but tetrahedrally coordinated V.\cite{PARIDA2019288} The lowest-energy structure of Ca$_2$VBiO$_6$ we found is $P2_1/c$, which is 168 meV/atom above the convex hull, while the experimental structure is $Cmc2_1$, which is a non-perovskite structure with VO$_4$ tetrahedral and (BiO$_2$)$^{-}$ chain.\cite{radosavljevic1998synthesis} This is presumably because V$^{5+}$ is too small for octahedral and therefore the tetrahedral coordination is more favorable than octahedral. Our calculated energy of Ca$_2$VBiO$_6$ with $Cmc2_1$ space group is 201 meV/atom lower than our $P2_1/c$ phase, which is more than enough to bring it to the convex hull. Ca$_2$AsBiO$_6$ is isostructural with Ca$_2$VBiO$_6$ and is expected to have a smaller energy difference because As$^{5+}$ is bigger than V$^{5+}$ and the octahedral coordinated As$^{5+}$ is less unfavorited. Our calculated $Cmc2_1$ Ca$_2$AsBiO$_6$ is 159 meV/atom lower in energy than $P2_1/c$, which is the lowest energy phase of our prototype search and is 159 meV/atom above the convex hull. Therefore, $Cmc2_1$ Ca$_2$AsBiO$_6$ is just on the convex hull of formation energy. Our relaxed structure of Ca$_2$PdWO$_6$ ($C2/m$), which is heavily distorted from $P2_1/c$ prototype structure because of the strong preference of Pd$^{2+}$ to square planer coordination rather than octahedral,\cite{doi:10.1021/acs.chemmater.7b02766} is 192 meV/atom lower in energy than the experimentally reported structure ($Pmm2$), where both W$^{6+}$ and Pd$^{2+}$ cations are octahedrally coordinated with six oxygen anions.\cite{Ca2PdWO6} However, our $C2/m$ structure is still 115 meV/atom above the convex hull, which means the experimental structure is 307 meV/atom above the convex hull. Therefore, it is plausible that the experimental structure is incorrect or the composition is not correct. Although La$_2$ZnTaO$_6$ is collected in the ICSD, the corresponding reference reported is actually La$_2$ZnTaO$_5$N.\cite{doi:10.1021/acs.inorgchem.1c00927} The experimentally synthesized La$_2$FeReO$_6$ and Sr$_2$MoZrO$_6$ are the $B$/$B^{\prime}$ disordered perovskite with space group $Pnma$.\cite{kaipamagalath2018griffiths,brixner1960x} The lowest-energy structure of our calculated Sr$_2$MoZrO$_6$ and La$_2$FeReO$_6$ double ($B$/$B^{\prime}$) perovskite is $P2_1/a$, which is 125 and 167 meV/atom above the convex hull, respectively. The configurational entropy induced by $B$/$B^{\prime}$ disorder is ln2 $K_{\rm B}$T ($K_{\rm B}$ is the Boltzmann constant and T is the temperature) and the free energy gain at 300 K is 17 meV/atom. Although these two compounds are metal and La$_2$FeReO$_6$ has magnetic moments and the electric and magnetic entropies have certain contribution to free energy as well, the gained free energy due to entropy is unlike to overcome such large energy (more than 100 meV/atom) at room temperature. Therefore, these two compounds are metastable. In summary, $\Delta{\rm E_{\rm stable}}$ = 0.1 eV/atom is a reasonable threadhold for metastable $A_2BB{^\prime}$O$_6$ compounds.

Since $A_2BB^{\prime}$O$_6$ are typical ionic compounds, the compounds that satisfy charge balance are indicated with open squares in the circles in Figure~\textcolor{blue}{S2} and \textcolor{blue}{S3} of the supporting information. The nominal charges of the elements studied in this work are from Pymatgen, see Table \textcolor{blue}{S2} of the supporting information. We can see that all compounds that are on the hull are charge-balanced, while most unstable compounds (\textit{i.e.}, $\Delta{E}_{\rm stab}$ $>$ 100 meV/atom) are not charge-balanced, which highlights the importance of charge balance in identifying stable/metastable $A_2BB{^\prime}$O$_6$ compounds. Therefore, charge balance can be used as a pre-screening criteria for perovskite discovery in the future. The full list of the stable and metastable compounds are tabulated in the supporting information csv table. As shown in Figure~\ref{basr} and~\ref{cala}, the $B$/$B^{\rm \prime}$ pairs of the stable/metastable Ba$_2BB^{\prime}$O$_6$ compounds are mainly from the combinations of 4$d$ and 3$d$ as well as 5$d$ and 3$d$ transition metals, but fewer pairs are from 4$d$ and 5$d$ elements. This is presumably due to the diverse oxidation states and smaller ionic radius of 3$d$ transition metal elements, which are easy to satisfy charge balance and favor octahedral coordination. Since Ba, Sr, and Ca are divalent cations, the elements with oxidation states from +1 to +7 can be $B$ and $B^{\prime}$, more charge-balanced compositions are expected in these divalent cations than trivalent La, where only the elements with oxidation states from +1 to +5 can be $B$ and $B^{\prime}$. This is inline with our observation of charge-balance in Figures \textcolor{blue}{S2} and \textcolor{blue}{S3} of the supporting information, as well as a previous study.\cite{vasala2015a2b} Lastly, many Technetium-containing compounds are predicted to be stable in this work, but they are not yet observed in experiments, which is likely due to its radioactivity.

\vspace{0.2cm}
\begin{center}
	\begin{table}
		\caption{{\bf Summary of the A$_2BB{^\prime}$O$_6$ compounds studied in this work. Stable (S), metastble (M), and unstable (U) are the compounds with convex hull distance of 0, 0 - 100, and $>$ 100 meV/atom, respectively. The experimentally observed compounds are under ``ICSD'' and ``Literature''. Here, ``Literature'' means the compounds are reported in literature but are not included in the ICSD.}}
		\begin{tabular}{c@{\hspace{1.7em}}c@{\hspace{1.7em}}c@{\hspace{1.7em}}c@{\hspace{1.7em}}c@{\hspace{1.7em}}c@{\hspace{1.7em}}c@{\hspace{1.7em}}c@{\hspace{1.7em}}c}
			\hline
			\hline
			$A$ site        &     S    &      M       & \multicolumn{3}{c}{ICSD}                   & \multicolumn{3}{c}{Literature}          \\
			\cmidrule(lr){4-6}  \cmidrule(lr){7-9}
			                &          &              &      S      &       M       &     U        &       S        &      M        &     U  \\
			\hline
			Ba              &  158     &     394      &      48      &     21       &     0        &       33       &      32       &     1  \\
			Sr              &  140     &     416      &      58      &     34       &     0        &       21       &      25       &     1  \\
			Ca              &   79     &     421      &      12      &     27       &     3        &       24       &      31       &     0  \\
			La              &   43     &     371      &      10      &     40       &     1        &       10       &      37       &     1  \\
			\hline									
		\end{tabular}
		\label{stablecomp}
	\end{table}
\end{center}

\vspace{0.5cm}
\noindent \textbf{Low-energy structure.}  As mentioned above, 10 prototype structures are employed in our high throughput DFT calculations to filter out the low-lying $A_2BB{^\prime}$O$_6$ compounds. Our DFT screening yields 381 stable double perovskites, 1404 metastable double perovskites, 38 stable non-perovskites, and 198 metastable non-perovskite compounds, which cover nearly all the compositions observed in experiments (see Table ~\ref{stablecomp}). The full lists of the perovskite and non-perovskite compounds are in single csv file of the supporting information.

\begin{figure}[th!]
	\setlength{\unitlength}{1cm}
	\includegraphics[width=1.0\linewidth,angle=0]{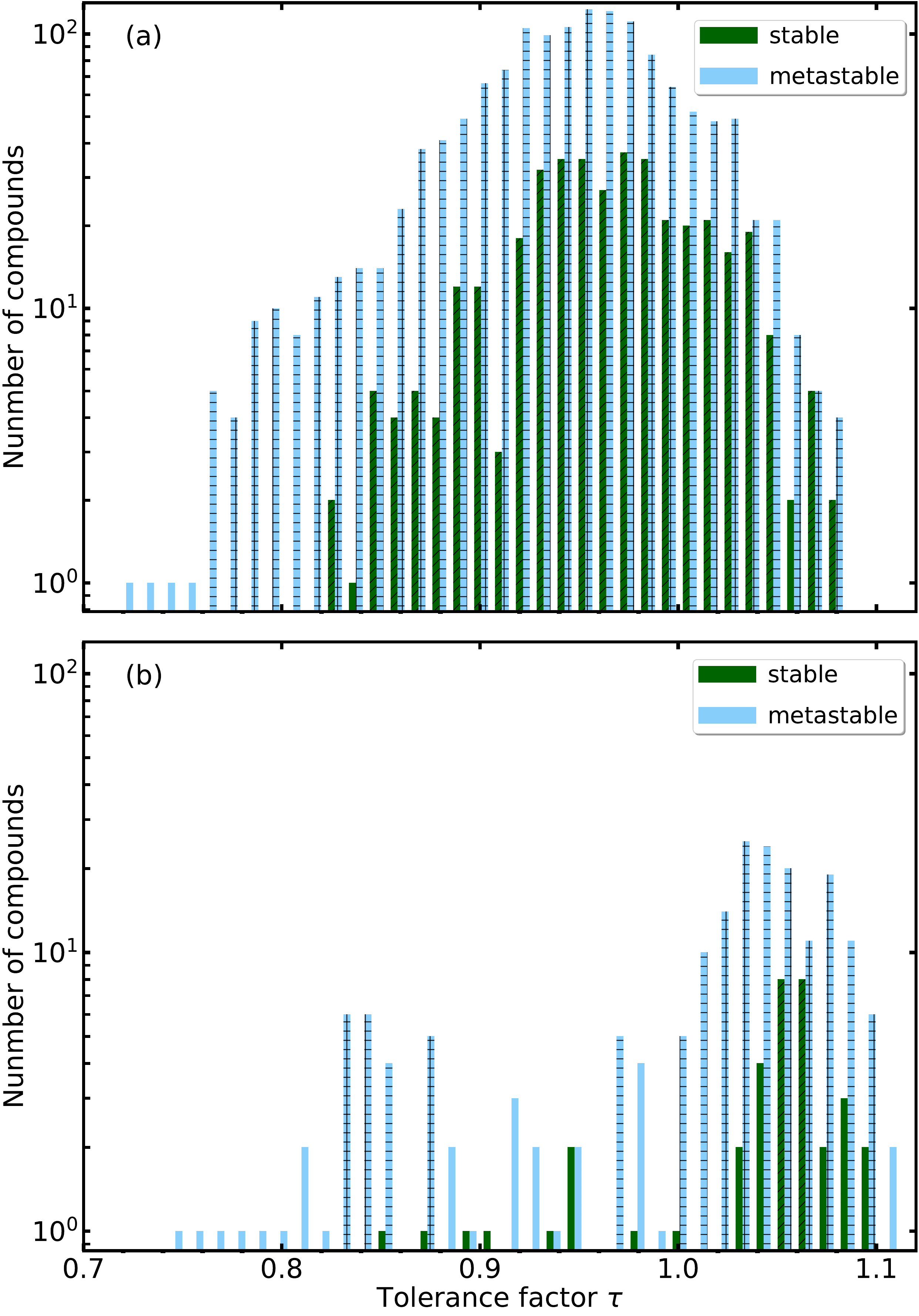}	
	\caption{The number of stable and metastable $A_2BB{^\prime}$O$_6$ double perovskite (a) and non-perovskite (b) compounds with Goldschmidt's tolerance factor.}
	\label{tolerance}
\end{figure}

Goldschmidt's tolerance factor ($\tau = \frac{r_{A} + r_{\rm O}}{\sqrt{2}(r_{B} + r_{\rm O})}$, where $r_{A}$, $r_{B}$, and $r_{\rm O}$ are the ionic radii of $A$, $B$, and O ions, respectively), which is an indicator of the mismatch between $A$-O and $B$-O layers, is widely used to determine the structural stability and distortion of perovskite.\cite{goldschmidt1926gesetze} The distribution of stable and metastable $A_2BB{^\prime}$O$_6$ perovskites with respect to Goldschmidt's tolerance factor is shown in Figure~\ref{tolerance}. The span of $\tau$ for stable double perovskite is from 0.82 to 1.10 and most of the stable double perovskites have $\tau$ around 0.97. The metastable double perovskite have a wider spread (from 0.72 to 1.11). This is expected because both small ($<$ 0.9) and large $\tau$ ($>$ 1.0) suggest large mismatch between $A$-O layer $B$-O layers, which result in large lattice strain and therefore low stability. Therefore, high-pressure and high-temperature methods are commonly used to synthesize perovskites and related compounds with low $\tau$.\cite{doi:10.1021/ja408931v,su2019predicted} Our finding is similar to the analysis for the perovskites observed in experiment.\cite{vasala2015a2b} 

\begin{figure}[htp!]
	\setlength{\unitlength}{1cm}
    \includegraphics[width=1.0\linewidth,angle=0]{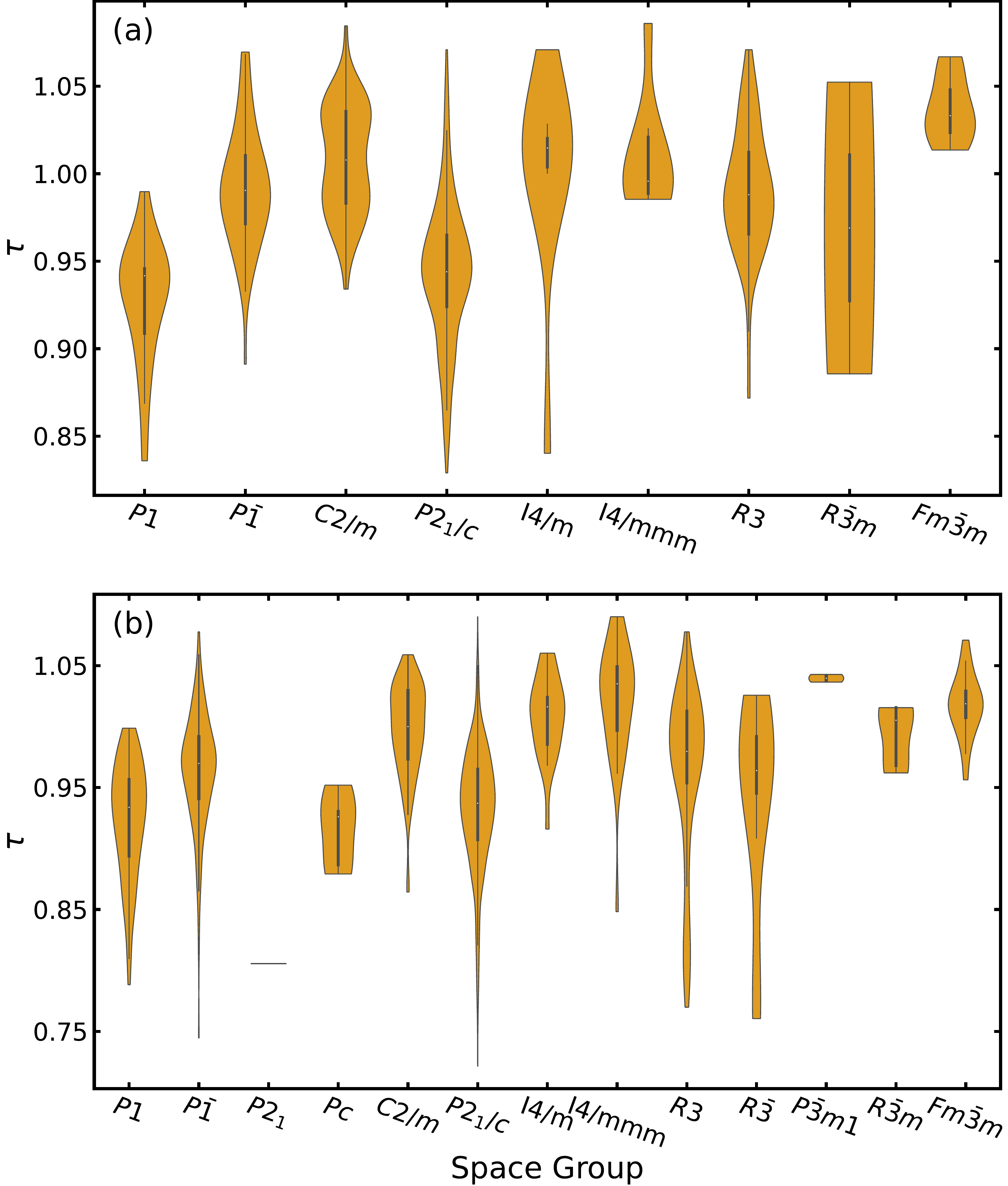}
	\caption{Distribution of the Goldschmidt's tolerance factor for stable (a) and metastable (b) $A_2BB{^\prime}$O$_6$ double perovskites with respect to space group.}
	\label{perovskitedist}
\end{figure}

The correlation between Goldschmidt's tolerance factor and space group of the stable and metastable perovskites is shown in Figure~\ref{perovskitedist}. Overall, the compounds with lower symmetry (lower space group number) has smaller $\tau$. Note there are more space groups than our prototype structures because of symmetry breaking during structure relaxation. For instance, $P1$ and $P\bar{1}$ are mainly from $P2_1/c$ and $C2/m$, respectively. As expected, $Fm\bar{3}m$ has $\tau$ $\sim$ 1.0 with a much narrow spread. This is slightly different from experiments because our calculations are performed at zero kelvin and the structure distortions are strongly dependent on temperature, \textit{i.e.}, the higher the temperature, the less the distortion.

\begin{figure}[htp!]
	\setlength{\unitlength}{1cm}
	\includegraphics[width=1.0\linewidth,angle=0]{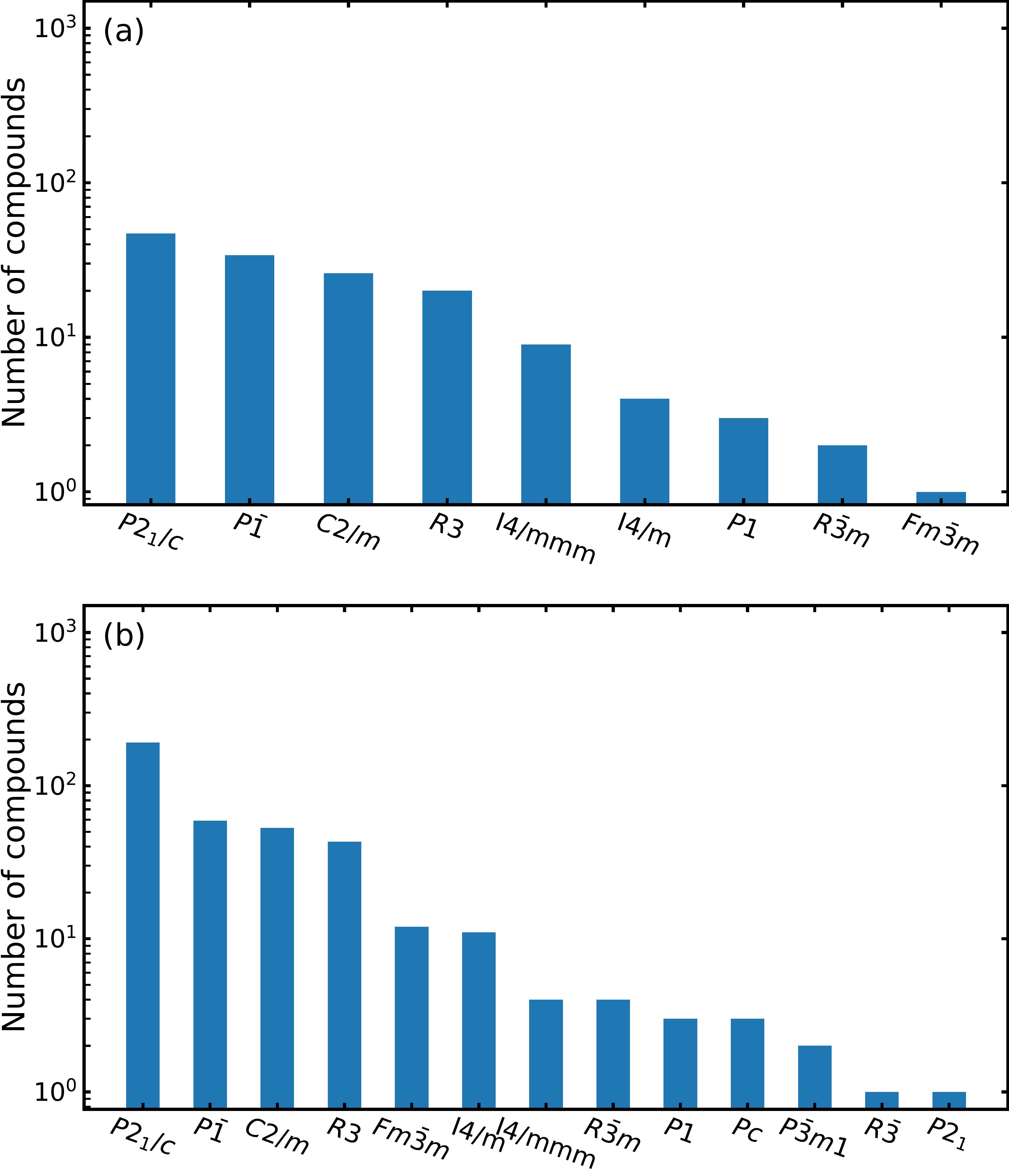}	
	\caption{Distribution of stable and metastable compounds for stable (a) and metastable (b) $A_2BB{^\prime}$O$_6$ double perovskites with respect to space group.}
	\label{all}
\end{figure}

The space group distribution of the stable and metastable double perovskites are shown in Figure~\ref{all}. Most of compounds adopt $P2_1/c$ space group, which is the combination of octahedral rotation and tilting, consisting with the experiment observation.\cite{vasala2015a2b} The second ($P\bar{1}$) and seventh ($P1$) favorable structures are not our prototype structures and these structures are due to the distortion during relaxation. Similarly, three space groups ($R\bar{3}$, $Pc$, $P2_1$) that are only observed in metastable compounds are due to symmetry breaking during structure relaxation. As previously mentioned, perovskites usually undergo structural phase transition associated with octahedral distortion and, the lower the temperature, the large octahedral distortion. Therefore, our 0 K calculations find more distorted perovskite than experiments.

Recently, Filip and Giustino revisited a rigid-sphere model of the perovskite geometry and proposed a generalised tolerance factor by taking into account electronegativity.\cite{doi:10.1073/pnas.1719179115} However, our results show that this tolerance factor has very similar results as Goldschmidt's tolerance factor, see Figure \textcolor{blue}{S4} and Figure \textcolor{blue}{S5} of the supporting information.
 
\vspace{0.5cm}
\noindent \textbf{Statistic analysis based on machine learning methods.} 
In this section, we use machine learning models to analysis formation energies and decomposition energies of our high throughput DFT calculated 4,855 compositions to unveil the hidden connection between formation energy (and decomposition energy) and composition. 431 features based on atomic characters of constituent elements are generated using Matminer\cite{WARD201860} and are adopted from literature. Twenty one most common machine learning algorithms are examined here, including Random Forest,\cite{breiman2001random} Gradient Boosting Regressor,\cite{friedman2001greedy} and XGBoost. In order to verify the performance of these models, the DFT datasets were shuffled and split into 80\% and 20\%  randomly for training and cross-validation, respectively. Note our training data is different from previous studies,\cite{doi:10.1021/acs.chemmater.7b00156,faber2016machine} we used the lowest energy of the 10 prototype structures screened from our high throughput DFT calculations for each composition, which takes into account different structure distortion and structural variability, and therefore is more challenging to build this model than the formation energies (and decomposition energies) of the compounds with the same structure, where the structural information is implicitly utilized in fact.

\vspace{0.5cm}
\noindent \textbf{Machine learning model of formation energy.}
As shown in Figure \textcolor{blue}{S6} of the supporting information, formation energies of these compounds are continuous and spread from -3.7 to -1.2 eV/atom with the maximum number of compounds at -2.5 eV/atom, which is appropriate to use regression models. Therefore, several commonly used regression models in material science are examined. The performance of all these models are presented in Figure \textcolor{blue}{S7} of the supporting information. Among them, XGBoost, Random Forest, and Gradient Boosting show the best performance. Therefore, we focus on these three models and carried out feature filter to avoid overfitting. We reduce the number of features from 589 to $\sim$ 30 by gradually removing the features that have less impact on model's performance, \textit{i.e.}, the feature has smallest effect on R$^2$ and MAE. All the generated features are employed in these 21 models and their performances are evaluated using R$^2$ and MAE. As we can see in Figure \textcolor{blue}{S8} of supporting information, the R$^2$ of Random Forest is not improved when the number of features is over 20, but it decreases rapidly when the number of features is less than 20. Therefore, we keep 20 features for the subsequent training in the Random Forest model. Similarly, we find 30 features are sufficient for Gradient Boosting. XGBoost, provides three feature sorting methods, namely gain, weight, and cover. The gain metric signifies the contribution of each feature during node splitting, and its calculation incorporates the reduction in loss resulting from feature splitting. If a feature can significantly reduce its loss when split, its gain will be higher. Weight represents the number of times the feature is used to split the node. During the construction of the tree, each feature is assigned importance based on the frequency with which it is used to split the dataset; the more times it is used, the higher its weight. Cover refers to the number of observations covered by a feature in all splits, and the coverage reflects the importance of the feature in the entire dataset. Since gain can quantify the contribution of each feature to loss reduction in each split, it directly measures the actual impact of each feature on model performance, thereby aiding in the understanding of the importance the model assigns to features. Therefore, the gain method is more robust compared to frequent splits and can help to reduce the risk of model overfitting. Then, we subsequently select the features obtained by gain as the input features of our XGBoost model. Results from weight and cover are presented as well, see Figure \textcolor{blue}{S9} of supporting information. Therefore, we choose 20 features with feature importance sorted by gain (labeled as XGBoost with gain), 30 features with feature importance sorted by weight (labeled as XGBoost with weight), and 35 features by cover (labeled as XGBoost with cover). Then, we conducted 20 training and cross-validation with randomly dataset splitting for each of these three models and ranked the feature importance after each training to obtain the most important features.

\begin{figure*}[htp!]
	\setlength{\unitlength}{1cm}
	\includegraphics[width=1.0\linewidth,angle=0]{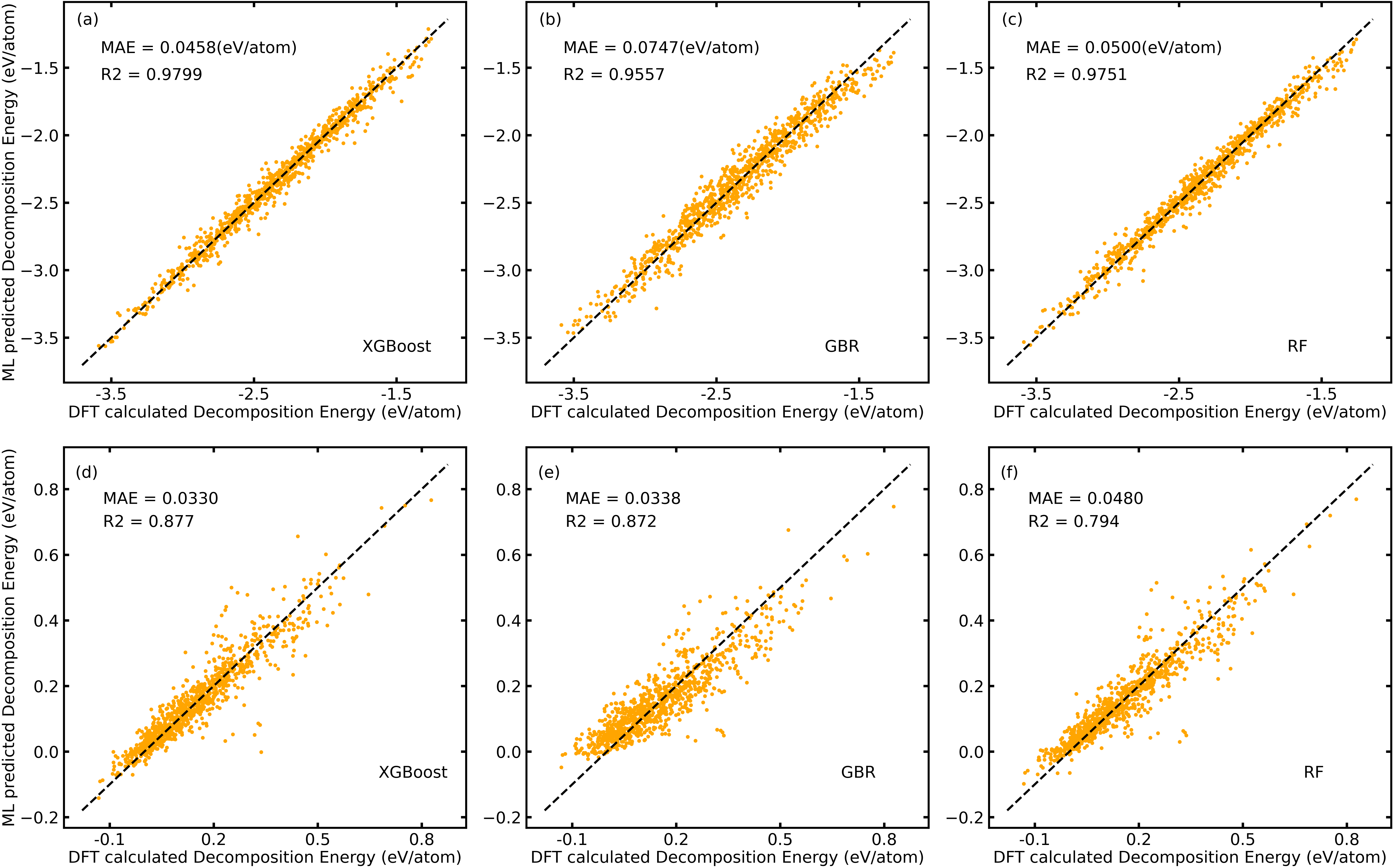}	
	\caption{Machine learning prediction vis DFT calculated formation energy and decomposition energy. (a)-(c) performances of XGBoost, GBR, and random forest (RF) algorithm on formation energy, respectively. (d)-(f)  performances of XGBoost, GBR, and Random Forest algorithm on convex hull distance, respectively.}
	\label{mlpredcit}
\end{figure*}

Then, we used Spearman's rank correlation to further filter the redundant and correlated features. For the top features from 20 training iterations for each model, we compute the Spearman's rank correlation coefficient between each pair of features, see Figure \textcolor{blue}{S10} of the supporting information. For any pair of features with a correlation coefficient greater than 0.8, we calculate the impact of the feature on the formation energy and remove the feature with lower correlation coefficient. This approach allows us to identify and eliminate non-independent features in a data-driven and statistically sound manner. After this feature filtering step, the number of features is further reduced for all these models. The final number of features for each model are: 14 features for Random Forest, 24 for Gradient Boosting, 20 for XGBoost with gain, 22 for XGBoost with weight, and 35 for XGBoost with cover. It is evident that Random Forest requires much less features than other algorithms. These reduced feature sets allow more efficient and accurate modeling formation energy while minimizing the risk of overfitting. Secondly, we can see that Random Forest, Gradient Boosting, and XGBoost models outperform others. XGBoost has the highest R$^2$ (0.9799) and smallest MAE (0.0458 eV/atom), which is followed by Random Forest (R$^2$=0.9751, MAE=0.0500 eV/atom) and Gradient Boosting (R$^2$=0.9557, MAE=0.0747 eV/atom), see Figure~\ref{mlpredcit}. However, the differences among these models are very small, indicating the robustness of our models. The results from other methods are presented in Table \textcolor{blue}{S3} and Table \textcolor{blue}{S4} of the supporting information. Our MAE is smaller than previous studies on perovskite oxides,\cite{faber2016machine,li2023center} especially considering we have more diverse crystal structures than these previous studies.

To unveil the underlying physics of our models, we conduct feature analysis for the three best models utilizing the SHAP algorithm,\cite{lundberg2017unified} which comprises two parts: the feature importance derived from the tree models themselves and the importance analysis generated by SHAP. The feature importance analysis shows that 6 of the top 10 features ranked by SHAP are common for all ML models, namely MagpieData mean Nunfilled (average number of unfilled electrons in the outermost shell of atoms in a compound), frac d valence electrons (Number of valence electrons in d orbital), avg ionic char (computational features in matminer for quantifying ionic properties in materials\cite{WARD201860}), MagpieData maximum Nvalence (Number of valence electrons), DemlData mean electronegativity, and DemlData maximum heat\_fusion (heat of fusion). Five features are common in at least two models, namely DemlData std\_dev atom\_radius (atomic radius), MagpieData maximum NdValence (the maximum number of valence electrons in the $d$ orbital of an atom in a compound), HOMO (the highest occupied molecular orbital) energy, frac $p$ valence electrons (the number of valence electrons in the $p$ orbital of an atom in a compound), and MagpieData range NdValence (the range of valence electrons in the $d$ orbital of an atom in a compound). Additionally, tree models also can obtain their own feature importance during training. We compare the SHAP feature importance analysis with that directly obtained  from these three tree models and find that the top-ranked features exhibit high similarity, \textit{i.e.}, the features related to the distribution of electrons outside the nucleus (MagpieData mean Nunfilled, frac d valence electrons, \textit{etc.}) have the highest rank. Secondly, properties such as electronegativity, thermodynamic properties (DemlData maximum heat\_fusion), atomic radius, and Mendeleev number have relatively lower rank. Although the ranks of these properties vary among different models, overall they are the properties correlated to electron distribution. Surprisingly, the tolerance factor, which plays important roles in structural distortion of perovskite, has low rank in formation energies of $A_2BB^{\prime}$O$_6$ compounds. More details of the feature rank can be found in Figure \textcolor{blue}{S11} of the supporting information.

\vspace{0.5cm}
\noindent \textbf{Machine learning model of stability.} 
As mentioned above, convex hull distance is defined as the energy difference between the formation energy of a compound with respect to the lowest formation energy of the linear combination of the competing phases with the same composition. Therefore, if the energy difference is less than zero, the convex hull distance is mathematically zero. However, regression models tend to yield more accurate results when the dataset exhibits a relatively even distribution across the range of values. In order to obtain a better ML model, decomposition energy, which measures the energy difference between the formation energy of a compound with respect to the lowest formation energy of the linear combination of the competing phases with the same composition, is adopted as the target here for stability prediction because it has a more even distribution than convex hull distance. As shown in Figure \textcolor{blue}{S6} of the supporting information, decomposition energy has a narrower spread (from -0.2 to 0.9 eV/atom) than the formation energy (from -3.7 to -1.2 eV/atom). Although it is expected that decomposition energy model would be worse than the formation energy one for regression with the same amount of data, the decomposition energy model is more useful in material discovery than the formation energy model. Our decomposition energy models are shown in Figure~\ref{mlpredcit}. As expected, all these models have lower R$^{2}$ and larger MAE than the formation energy models. Similar to the results of formation energy models, XGBoost has the best performance (high R$^2$ and small MAE) among these three models, which is followed by Random Forest and gradient boosting regression. Overall, the performance of our model surpasses or is on par with that of previous studies.\cite{doi:10.1021/acs.chemmater.7b00156,LI2018454,https://doi.org/10.1002/adfm.201807280}

\begin{figure}[th!]
	\setlength{\unitlength}{1cm}
	\includegraphics[width=1.0\linewidth,angle=0]{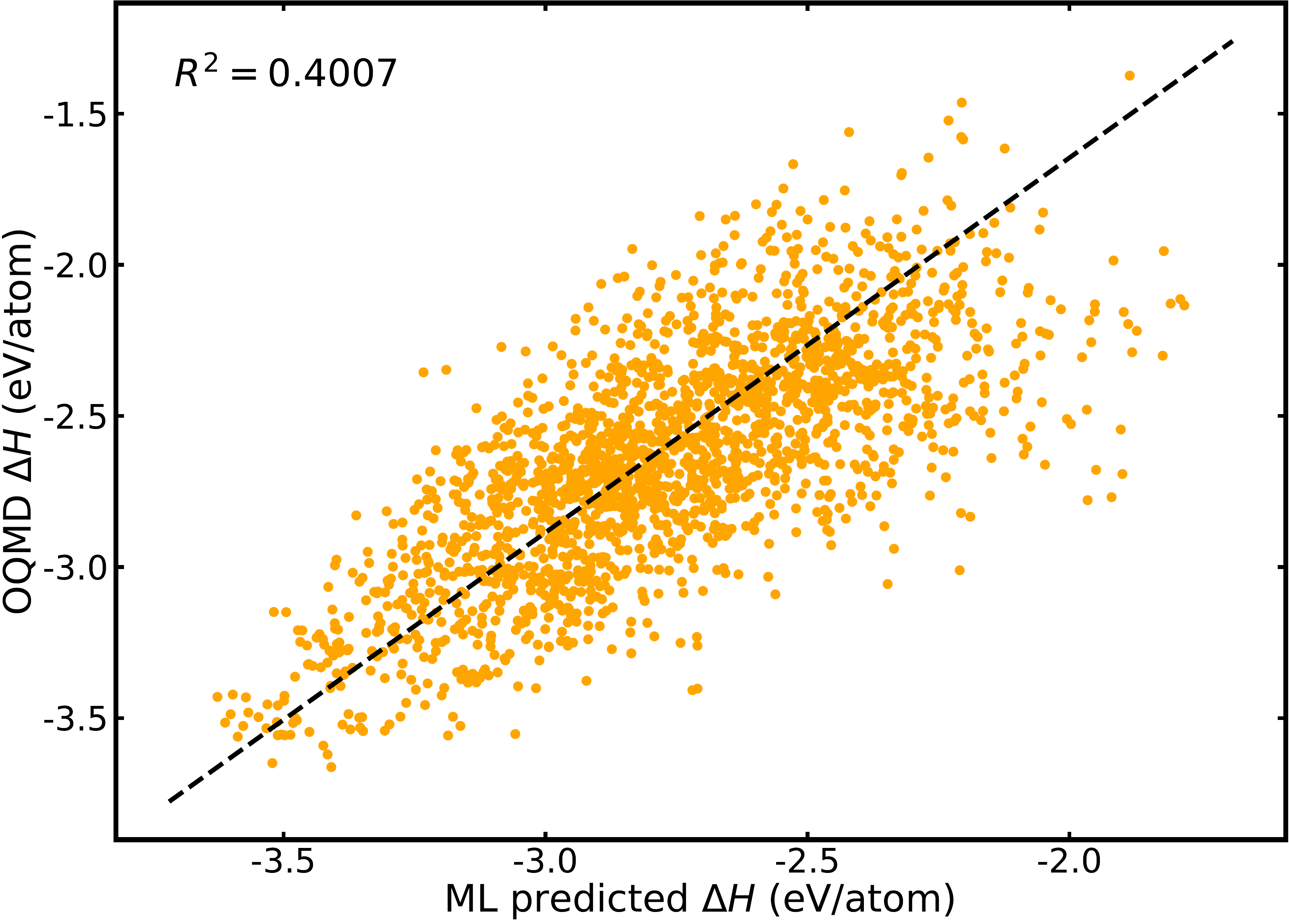}	
	\caption{{Comparison of XGBoost prediction results for formation energy with OQMD database results.}}
	\label{mloqmd}
\end{figure}

\vspace{0.5cm}
\noindent \textbf{Application of machine learning model for new material discovery.} The power of machine learning lies in its ability to expedite material discovery by directly predicting the thermal stabilities of a composition without prior knowledge of its structure. Subsequent computations only are needed for the compositions that are predicted to be stable, which can save a lot of CPU hours. Since we have calculated all possible $A_2BB^{\prime}$O$_6$ compositions within the chemical space $A$= Ca, Sr, Ba, and La, we then use our ML models to predict the stability (decomposition energy) of the $A_2BB^{\prime}$O$_6$ compositions that are not included in our dataset, namely the $A_2BB^{\prime}$O$_6$ compounds with $A$ = Ce, Pr, Nd, Sm, Eu, Gd, Tb, Dy, Ho, Er, Tm, Yb, and Lu and same $B$ and $B^{\prime}$. We compare our predicted stabilities of these compounds with those that already exist in the OQMD. Although these elements are not included in our training set, our ML models are based on the features of constituent elements and should be extendable to these elements. We screen 2071 compositions and predict their formation energy and decomposition energy using our XGBoost model. The prediction results of the formation energy are shown in Figure~\ref{mloqmd}, the results predicted by XGBoost are positively correlated with the results of the OQMD. For decomposition energy, 803 compositions are predicted to be stable or metastable (with decomposition energy $\leqslant$ 0.1 eV/atom) and 612 of them have $\Delta{\rm stable}$ less than 0.1 eV in the OQMD, indicating a 76.2 \% success rate.

\vspace{0.5cm}
\noindent $\blacksquare$ \textbf{CONCLUSIONS} \\
In summary, we explored the thermal stabilities of $A_2BB^{\prime}$O$_6$ compounds ($A$=Ca, Sr, Ba, and La; $B/B^{\prime}$ are metal elements), using high throughput DFT calculations. We not only rediscovered 470 compounds that have been reported in the ICSD and literature, but also discovered 1,552 new synthesizable $A_2BB^{\prime}$O$_6$ compounds. Our analysis shows that most $A_2BB^{\prime}$O$_6$ compounds adopt $P2_1/c$ structure at 0 K, which is consistent with tolerance factor analysis. Using 4,855 formation energies and decomposition energies from our high throughput DFT calculations, we build several machine learning models based on the element features and ionic radius of constitution elements. The MAE of formation energy and stability are as low as 0.0444 and 0.0330 eV/atom, respectively. Formation energy and stability are strongly correlated with the number of empty electron of constitution elements. We then used our stability model to predict new stable/metastable $A_2BB^{\prime}$O$_6$ compounds that are outside the chemical space explored in the high-throughput study. A success rate of 76.2 \% is achieved by comparing our ML prediction with those that already exist in the OQMD. Our work delineates the stability map of $A_2BB^{\prime}$O$_6$ compounds and offers new insights in $A_2BB^{\prime}$O$_6$ compounds discovery.

\vspace{0.5cm}
\begin{suppinfo}
The Supporting Information is available free of charge on the ACS Publications website at DOI: \\
The prototype structures; the convex hull distance of the charge-balanced compounds; the tolerance factor of $A_2BB^{\prime}$O$_6$; the performance of all the machine learning models used in this work; feature importance of the selected models; the Spearman correlation coefficient of the selected features; models and parameters used in this work.
\end{suppinfo}

\vspace{0.5cm}

\noindent $\blacksquare$ \textbf{AUTHOR INFORMATION} \\
\noindent \textbf{Corresponding Author} \\
*(C.W.) E-mail: c-wolverton@northwestern.edu \\
*(J.H.) E-mail: jghe2021@ustb.edu.cn \\

\vspace{0.5cm}
\noindent \textbf{Notes} \\
The authors declare no competing financial interest.\\

\vspace{0.5cm}
\noindent $\blacksquare$ \textbf{ACKNOWLEDGMENT}
Y.W. and J.H. acknowledge the support of the Fundamental Research Funds for the Central Universities China (USTB).
C.W. J.S., and B.B. acknowledge primary support via National Science Foundation through the MRSEC program (NSF-DMR 1720139) at the Materials Research Center.
\\

\bibliography{ref}

\begin{tocentry}
\includegraphics[clip,width=0.95\linewidth]{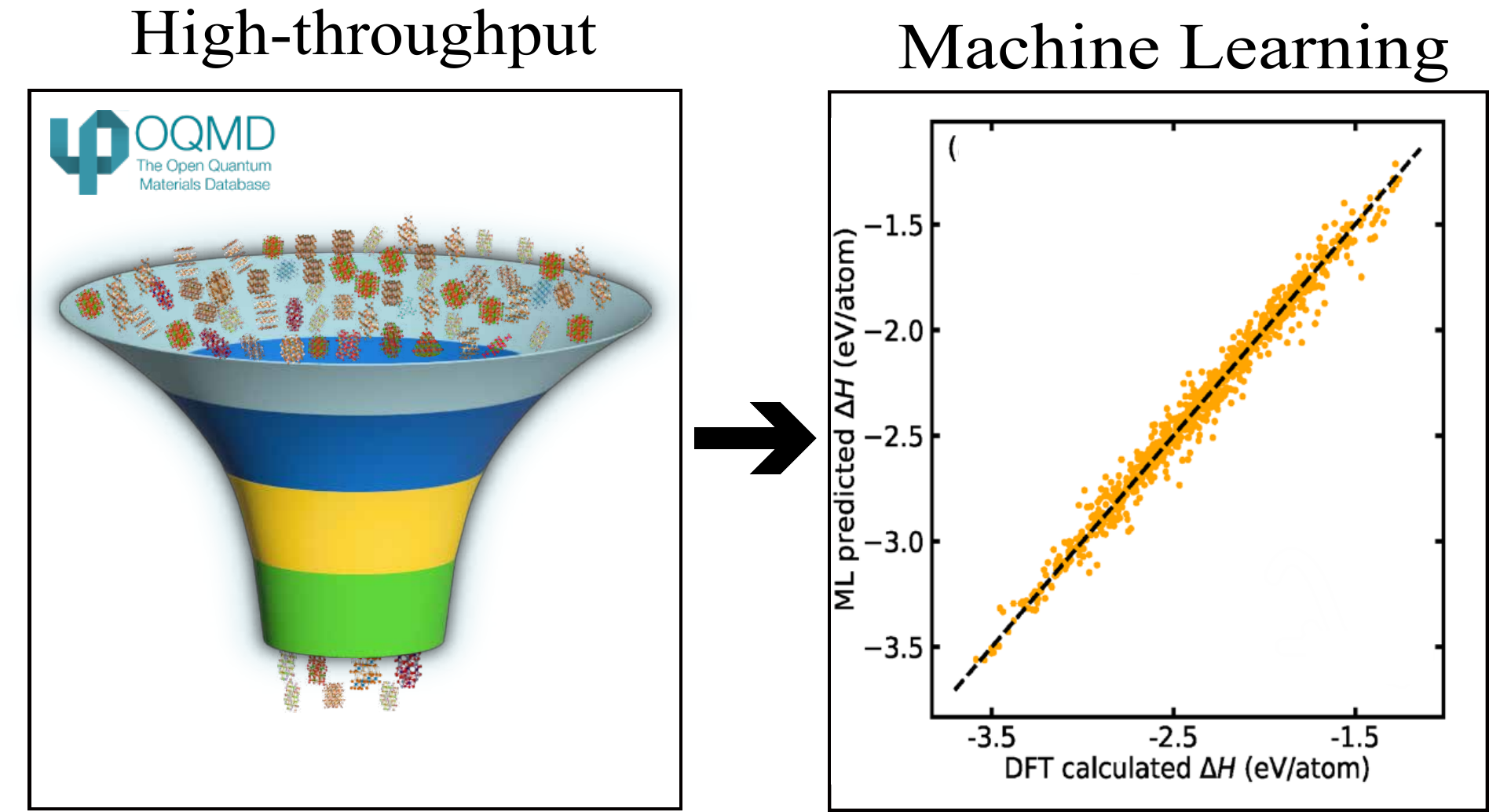}
\end{tocentry}

\end{document}